\documentclass[12pt]{iopart}
\eqnobysec
\def\eqn{\begin{equation}}
\def\enq{\end{equation}}
\def\ena{\begin{array}}
\def\eqa{\end{array}}
\def\CM{{\cal M}}
\def\CN{{\cal N}}
\def\CW{{\cal W}}
\def\sLm{sL-submanifold} 

\def\Im{{\rm Im\ }}
\def\bra#1{{\langle}#1|}

\def\II{II}

\def\IIb{IIb}
\def\p{\partial}
\def\tr{{\rm tr\ }}
\def\Tr{{\rm Tr\ }}
\def\half{{1\over 2}}
\def\BC{{\bf C}}
\def\BP{{\bf P}}
\def\BR{{\bf R}}
\def\BZ{{\bf Z}}
\begin{document}
\title{Topics in D-Geometry}
\author{Michael R. Douglas}
\address{
Department of Physics and Astronomy\\
Rutgers University\\
Piscataway, NJ 08855--0849; USA}
\address{{\it and}}
\address{
I.H.E.S., Le Bois-Marie, Bures-sur-Yvette, 91440 France}
\begin{abstract}
We discuss the general theory of D-branes on Calabi-Yaus, recent results
from the theory of boundary states, and new results on the spectrum of
branes on the quintic CY.
(Contribution to the proceedings of Strings '99 in Potsdam, Germany.)
\end{abstract}
\section{Introduction}

The present contribution consists of three parts.
The first is a general summary of the theory of D-branes on Calabi-Yau;
the second summarizes the works \cite{bdlr,dg}\ which connect
the boundary state approach with large volume results;
the third summarizes new results on lines of marginal stability on the
quintic found in June 1999.
The transparencies for this talk (which emphasize different parts
of the material) are also available at \cite{trans}.

For background material on ``D-geometry,'' see \cite{doug}.  This term refers
to the study of how the conventional geometry
which describes branes in supergravity
is generalized in the context of D-branes.
As a point of departure we could consider any of the
geometrical pictures which branes give us for the various terms in
an effective action.  Perhaps the simplest example is the following:
the moduli space of a $0$-brane at a point in a CY$_3$
is the CY$_3$ itself; the moduli space metric is just the Ricci-flat
metric on the CY$_3$.

Examples of the ``unconventional'' geometry we have in mind include the
following:
\begin{enumerate}

\item
Stringy and quantum corrections will generally modify conventional geometric
predictions.  In particular, we can ask how a D-brane world volume
action is affected by``stringy'' ($l_s$) corrections.  An example is
to find the moduli space metric for the D$0$-brane at a point; this
provides a canonical non-Ricci flat metric for each point in CY moduli space.
Qualitative effects visible at finite $l_s$ include T-duality and mirror
symmetry; we will discuss the latter below.

\item Perturbative string compactification can be defined
non-geometrically, by specifying an appropriate internal CFT.  Some
examples (such as Gepner models) turn out to have geometric interpretations,
and this definition provides a concrete way to work in the ``highly
stringy'' regime.  Others such as asymmetric orbifolds do not have
known geometric interpretations; studying D-branes on these spaces will
probably lead either to finding such interpretations or showing why they
do not exist.

\item D-brane world-volume theories include open strings stretching
between pairs of branes, which in many cases provide alternate gauge
theory origins for what are gravitational effects in the large distance
limit.  Orbifold resolution by quiver theories are an example in which
non-trivial topology is reproduced as a classical gauge theory moduli space.
The short distance gravitational interactions between D-branes are
replaced by quantum gauge theory dynamics.  In special
cases (in the large $N$ limit or for quantities protected by supersymmetry)
this is believed to reproduce supergravity, but more generally provides another
way of defining its stringy generalization.

\item Noncommutative gauge theory arises on D-brane world-volumes
in appropriate limits of string theory, such as compactification on
a small torus with fixed background $B$ field, or in Minkowski space
with large $B$ field.  It seems quite likely that similar theories are
relevant in curved backgrounds; finding concrete examples is an important
problem for future work.

\end{enumerate}

This is by no means a complete list but perhaps includes the most interesting
points
discovered so far.  As each of them would form a topic in its own right, for
the
rest of the review we will focus on the following meta-question: to what extent
do
these effects lead to {\it qualitative} changes in the brane physics -- and
thus
cannot be ignored?  The way to study this question is to frame the alternative
(null) hypothesis: the qualitative properties of brane theories (especially,
the
low energy effective action, dimension of the moduli space, types of
singularities
and so on) are the same as predicted by naive geometric considerations -- and
test
it in examples.  We will refer to this as the ``geometric hypothesis'' and make
it
more precise below.

\section{D-branes on Calabi-Yaus}

Quite a lot is known about D-branes in flat space (Minkowski or
toroidal compactifications) and in K3 compactifications,  where type
\II-heterotic duality and the large supersymmetry already suffice to
give a good picture.  The geometric hypothesis appears to be essentially true
in
these cases --
the brane spectrum and moduli spaces can be described as the spectrum
and moduli spaces of semistable coherent sheaves (a generalization of vector
bundle which allows singularities corresponding to pointlike instantons)
\cite{hm}.

D-branes on Calabi-Yau threefolds are not so well understood and look
quite interesting for a number of reasons.  Physically, supersymmetry
preserving branes will have $\CN=1$, $d=4$ gauge theories on the world-volume
which may be directly relevant for phenomenology.
They generalize the strong coupling limit of heterotic string compactification
but in some ways appear simpler than the $(0,2)$ sigma models which appear
there.  Many questions can be addressed using the highly developed theory
of $\CN=2$ supersymmetry and mirror symmetry.

An important difference with the cases of higher supersymmetry is that the
spectrum of branes can depend on the particular vacuum (point in moduli space)
under discussion.  For example, in pure
$SU(2)$ gauge theory, we know that the strong coupling spectrum is quite
different
from the semiclassical spectrum; the purely electric ``W bosons''
are not present.  Given $\CN=2$ supersymmetry
this dependence of spectrum on moduli is
highly constrained: as is well known, the BPS
spectrum can change only on lines of marginal
stability defined by the condition $\Im Z(Q_1)/Z(Q_2)=0$.
Thus the problem of finding the spectrum of wrapped branes on CY and
deciding whether it too changes at string scales is non-trivial but
accessible, as we will discuss in the next sections.

Supersymmetric ($1/2$ BPS) branes on a CY$_3$ are divided into A and B branes
depending on the boundary condition on the $U(1)$ currents in the
$(2,2)$ superconformal algebra (which determines
which part of the world-sheet supersymmetry they preserve) \cite{ooy}:
either $Q_L = +Q_R$ or $Q_L=-Q_R$ is a consistent choice.
The notation comes from topological field theory -- an A brane is one
whose open strings naturally couple
to A-twisted topological theory and the K\"ahler
moduli, while a B brane couples to complex structure moduli.
Mirror symmetry will exchange the two -- the spectrum and world-sheet
theories of A branes on a CY $\CM$ is isomorphic to that of the B branes
on its mirror $\CW$.

If we consider branes defined by Dirichlet and Neumann boundary conditions
in the non-linear sigma model with CY$_3$ target,
the B branes are $2p$-branes wrapped on
holomorphic cycles and carrying holomorphic vector bundles
(this is the case with direct analogy to the heterotic string), while
the A branes are $3$-branes wrapped on what are called special
Lagrangian submanifolds (or sL-submanifolds; more below) \cite{bbs}.
At first this notation may seem backwards given the discussion in the previous
paragraph, since the $2p$-cycles and the masses of B branes are controlled
by K\"ahler moduli (and thus are calculable in the A-twisted topological closed
string theory), while the $3$-cycles and masses of A branes are
controlled by complex structure moduli.
Nevertheless it is correct  -- in going
from the open to closed string channel the boundary conditions on
the $U(1)$ current change sign, interchanging A and B twistings.

This switch has important consequences, especially if we combine it
with the known properties of CY sigma models.  Specifically, the B twisted
models receive no quantum corrections, while A twisted models receive
world-sheet
instanton corrections. Physically, this means that the
$\CN=2$ prepotential in compactified \IIb\ theory, which depends only
on complex structure moduli, is classically exact.
This means that whereas B brane masses receive world-sheet instanton
corrections,
the large volume results for central charges and masses of A branes are already
exact
(this fact and mirror symmetry can then be used to determine B masses).

This means that lines of marginal stability for A branes are the same as in the
large volume limit, and this fact strongly suggests that
the spectrum of A branes is determined entirely by
classical geometric considerations.  Since we have not argued that
the world-volume theory itself does not receive stringy corrections
(indeed we expect it to), this might seem to be an unjustified leap of faith
at this point.  Nevertheless there is a good argument for it, which we now
summarize.

The classical geometric prediction is that each A brane is a $3$-brane wrapped
on a {\sLm}.
Now a \sLm\ $\Sigma$ of a CY $n$-fold
is a Lagrangian submanifold with respect
to the K\"ahler form: $\omega|_\Sigma=0$, satisfying an
additional constraint involving the holomorphic $n$-form:
there exists a constant $\theta$ such that
\eqn
\Im e^{i\theta} \Omega|_\Sigma = 0.
\enq
The constant $\theta$ determines which of the original $\CN=2$ supersymmetries
remains unbroken; two branes of different $\theta$ together break all
supersymmetry.

While Lagrangian submanifolds are ``floppy,'' specified locally by
an arbitrary function (in canonical coordinates, $p_i=\p f/\p x^i$),
the special Lagrangian condition determines this function up to a finite
dimensional moduli space, which for a smooth CY has been shown to be
smooth and of real dimension $b^1=\dim H^1(\Sigma,\BR)$ \cite{mclean}.
A D-brane configuration
is specified by $\Sigma$ and a flat $U(1)$ gauge connection, leading to
a moduli space of complex dimension $b^1$, which before
taking stringy corrections into account is a torus fibration.

Interesting examples of {\sLm}s of $\BR^6$ are known,
but not too many are known for CY's.
The only general construction known is as the fixed point of
an involution, i.e. $\Im z^i=0$ in a CICY.
Even necessary or sufficient conditions for candidate cycles to support
{\sLm}s are not known.
The subject is still rather new however and interest
has picked up dramatically as a consequence of the
proposal of Strominger, Yau and Zaslow that the mirror $\CW$ to a CY $\CM$
is just the moduli space of the D$3$-brane on $\CM$
mirror to the D$0$ on $\CW$,
which will be some (appropriately chosen) $T^3$. \cite{syz}
A number of papers have shown the existence of $T^3$ fibrations
on particular CY's
which can in principle be deformed to special Lagrangian fibrations.
\cite{grossruan}

The question of how deformations of the CY itself affect the spectrum
of {\sLm}s has recently been studied by Joyce.
\cite{joyce}
The part of this story relevant for complex structure deformations
(also summarized in \cite{kachru}) is as follows.

The natural geometric description of transitions between
$3$-brane configurations in six dimensions is for two intersecting $3$-branes
to intercommute, producing a single $3$-brane, or the reverse.
In the large volume limit,
this process can be studied in the neighborhood of the intersection point,
and the relevant question is: out of all
configurations $\Sigma_\Theta$
in $\BR^6$ which asymptote to two planes $\Sigma_1$ and $\Sigma_2$
at fixed angles $\Theta$, is the minimal volume surface the union of the two
planes, or something else, and if so what?

This question was answered some years ago by use of calibrated geometry
\cite{harvey}\ and the result is known as the ``angle theorem'':
let $\Sigma_1$ be the first plane and $\bar\Sigma_2$ the orientation
reversal of the second plane; out of $SO(2n)$ rotations turning
$\Sigma_1$ into $\bar\Sigma_2$ take the eigenvalues $e^{i\theta_i}$
and let $\theta=\sum \theta_i$.
If the minimal such $\theta$ is greater than or equal
to $\pi$, the volume cannot be reduced; while if $\theta<\pi$ it can.

The surface of lower volume can be approximately described by use of an exact
\sLm\ solution in $\BR^6$ with the prescribed asymptotics, which exists in the
case
$\theta=\pi$.  One can try use this solution to lower the volume by orienting
it to cross both of $\Sigma_1$ and $\Sigma_2$
near the intersection point; if it does so, the finite region
between the intersections is guaranteed to have lower volume
than the original planes.  This will be possible exactly when $\theta<\pi$.

The angle theorem tells us which of two configurations is stable in terms of
a local geometric condition (the same as the string theory condition
for the intersection point to have an associated tachyon \cite{bdl}), but
the geometric picture furthermore implies that this
can be tested just knowing the central charges for the two branes.
This is because the relative angle is known given the phase of
pullback of $\Omega$ (locally $dz^1\wedge dz^2\wedge dz^3$) to each brane, and
$\Omega$ must have constant phase on each brane.
Thus decays take place just when $Z(Q_1)$ and $Z(Q_2)$ are colinear --
this is exactly the standard marginal stability condition.
These considerations tell us a little more -- namely, which state (the single
brane, or two branes) is stable on which side of the marginal stability line.

This geometrical picture
of A brane decay and stability fits with the constraints following from the
exact stringy
prepotential and thus, despite the
fact that other consequences of this geometrical picture may well be false
for substringy branes, it is consistent to imagine that the spectrum is the
geometric
one.  This is in contrast to the B description of the spectrum
which must be modified by the stringy
corrections to the prepotential.  This is the first example of what we will
call
below the ``modified geometric hypothesis.''

All of this tells us quite a bit about the dependence
of the spectrum of $3$-branes on the CY moduli, but does not substitute for
the need to have some results on the spectrum in at least some part
of moduli space.  Since so little is known about $3$-branes at present we
instead take this from the large volume limit of the
B brane spectrum as many mathematical results
towards classifying holomorphic cycles and vector bundles are known.

The most basic of these is the following.
Given a holomorphic vector bundle,
the Donaldson-Uhlenbeck-Yau theorem gives necessary and sufficient conditions for the
existence of a Yang-Mills connection preserving supersymmetry: it must
be semistable.  This is a somewhat complicated condition involving all holomorphic
subbundles, but a simpler necessary condition is known which depends on
the Chern character of the bundle (which corresponds to D-brane charge
as $Q_{6-2k}\equiv ch_k(F)$, the $2k$ form in $\Tr e^{F/2\pi}$) and the K\"ahler
class:
\eqn\label{stability}
\int (Q_6 Q_2 + \half Q_4^2) \wedge \omega \ge 0.
\enq
On manifolds with $b^{1,1}>1$ this describes an
explicit dependence of the spectrum on the K\"ahler class,
as has been discussed by Sharpe. \cite{sharpe}

Since the prepotential determining the central charges of B branes
receives world-sheet instanton corrections, it is fairly certain
that this mathematical stability condition is modified in the stringy
regime.  This is quite interesting as it would mean that the condition for
a bundle to be usable in superstring compactification is {\it not} always the
geometrical condition which has been implicitly assumed in the past.

Given a specific supersymmetric brane, we can try to derive its world-volume
effective action and general considerations suggest that the simplest
quantities to start with are the holomorphic ones: the superpotential
and gauge kinetic term.  The latter corresponds to the dilaton and
in CY compactifications with zero NS field strength this only becomes
non-trivial
at string loop level (this is one of the invariants
defined in \cite{bcov}).
However a superpotential can appear at tree level and indeed for multiple
parallel branes we expect a generalization of the $\tr Z^1[Z^2,Z^3]$
superpotential of $3$-branes in flat space.
There are also known examples of superpotentials for single branes
(see \cite{bdlr}\ for a discussion).

A plausible counterpart of the nonrenormalization theorem for the
$\CN=2$ prepotential is the following: the superpotential, being
essentially a topological quantity in open string theory, depends
only on the moduli of the appropriate twisted theory. \cite{bdlr}
Specifically,
an A brane superpotential depends only on K\"ahler moduli, while a
B brane superpotential depends only on complex structure moduli, and
furthermore is equal to the large volume result.

This comes close to showing that a B brane moduli space is the same as in
the large volume limit, but not quite -- the potential can also contain D
terms.
These would naturally depend on the K\"ahler moduli, as in the example
of quiver theories.  A natural generalization of the preceding conjecture is
that
these could be determined in the large volume limit from the A brane point of
view.

As explained in \cite{kachru}, the D terms are related to the stability
question.
A world-volume description of the decay process of Joyce starts with the two
intersecting 3-branes and $U(1)\times U(1)$ gauge theory; the intersection
comes
with a chiral multiplet charged under both $U(1)$'s, and the dependence on
complex structure moduli comes through an FI term for the relative $U(1)$.  As
one goes through the transition one goes from a supersymmetry breaking ground
state with unbroken $U(1)\times U(1)$ to a supersymmetry preserving ground
state with broken relative $U(1)$.  \footnote{
As in \cite{calmal} this configuration can be shown
by a probe analysis to be equivalent
to a single $3$-brane produced by intercommutation. \cite{dta}
A difference with the case of $2$-branes
considered there is that the resulting $3$-brane is not actually
special Lagrangian.  This is possible because near the transition
it has a large extrinsic curvature and this is evidence
that the detailed form of the special Lagrangian condition gets $\alpha'$
corrections.}

An analogous statement was already known on the B side. \cite{sharpe}
Equality in (\ref{stability}) defines a boundary within the K\"ahler cone
on which stability degenerates to semistability.  This means that the connection 
on the brane becomes reducible, and an enhanced gauge symmetry appears,
a phenomenon which in $\CN=1$ theory can only arise from D terms as above.
We see that this qualitative picture survives the stringy corrections, 
but the precise location of the boundary is different, in a way determined
by the A picture geometry.
\smallskip

The upshot of the discussion is that mirror symmetry leads to a natural
conjecture for a modified or ``mirror geometric hypothesis'' -- some brane
questions
are geometric in the A picture, and others are geometric in the B picture.
As is well known the prepotential in the complex structure sector is
determined geometrically; this determines A brane central charges and stability
and strongly motivates the claim that the spectrum of branes can be understood
geometrically in the A picture.
We can add to this the claim that the superpotential in the B twisted model
is classical; this means that brane moduli spaces are largely
determined by the geometry of the B picture.  Finally, it may be possible
to determine the D terms in the A picture and complete the story.

So far as I know, these conjectures are consistent with the evidence, but
require
much more testing.  The most interesting tests are in the stringy regime,
as we discuss next.

\section{Boundary states and branes}

Exactly solvable CFT's were a fruitful source
of insight into compactification of closed string theory and are now
beginning to teach us about branes in these compactifications.
The fundamental notion is that of ``boundary state,'' a CFT description
of a boundary condition as a linear functional on the closed string
Hilbert space.  Reparameterization invariance and supersymmetry can
be easily implemented by imposing operator constraints.  One must then
impose the condition that all annulus partition functions
(associated with pairs of boundary states) have an open string Hilbert space
interpretation (the multiplicities are integers); this condition was
proposed by Cardy and can be solved for rational CFT's.
D-brane ground states correspond to such boundary states
(not much is known about the non-rational case; possibly
additional unknown constraints must be satisfied).

The simplest and most studied models are orbifolds and orientifolds.
In this case the general boundary state approach
can be shown to reduce to the world-sheet prescription proposed in
\cite{dm} -- one introduces image D-branes on the cover and quotients
by a simultaneous space-time and gauge action.
The case of strings and branes near a single orbifold or orientifold
singularity is particularly easy and one obtains quiver gauge theories
as world-volume theories.
For $\BC^3/\Gamma$
these have been much studied and among the noteworthy results are the
following:
\begin{enumerate}
\item  The resolution of these singularities is described in
quiver gauge theory
by FI terms coupling to K\"ahler moduli. \cite{dgm,sardoinfirri}
If multiple resolutions with different topology are mathematically
possible, they all appear to be accessible physically. \cite{gretc}
\item The resulting metrics are not Ricci flat. \cite{douggreene,sardoinfirri}
Although some caveats were made in that work, it can be
shown that this statement is true at string tree level. \cite{dta}
\item The quiver theory depends on the choice of representation of $\Gamma$;
the basic case is the regular representation, while non-regular representations
correspond to branes wrapped around exceptional cycles (or ``fractional
branes'').
\cite{fract}
\item If we take D$3$-branes to get a $3+1$ theory, the regular representation
is distinguished by having zero beta function in the
large $N$ limit.\cite{lnv}
\item These theories have supergravity duals corresponding to the quotients
$AdS_5 \times S^5/\Gamma$.\cite{ks}
\end{enumerate}

Recently Diaconescu and Gomis have studied the case of $\BC^3/\BZ_3$ in
detail.\cite{dg}
Besides checking the equivalence between the boundary state approach
and the proposal of \cite{dm}, they determined the mapping between fractional
branes and wrapped branes in the large volume limit, using techniques we will
describe below.  Additional summary of this example can be found in
\cite{trans}.

\smallskip

We now turn to Gepner models and the work \cite{bdlr}.
Gepner models provide CFT models which are equivalent to CY compactification
at special points in moduli space of enhanced discrete symmetry.
The study of boundary states in these models was initiated by Recknagel and
Schomerus \cite{rs};
they classified the subset of boundary states which can be obtained by separate
boundary conditions in the individual $\CN=2$ minimal model factors, for which
Cardy's techniques apply.
(See also \cite{gutperle}, as well as
\cite{govind} which uses the Landau-Ginsburg approach.)

Let us briefly summarize the spectrum of branes one obtains and the main result
used in the analysis of \cite{bdlr} -- the intersection form between two
branes.
Cardy's analysis (for diagonal modular invariant)
produces boundary conditions in one-to-one correspondance with
closed string primary fields;
the spectrum of open strings with two such boundary conditions $a$ and $b$
is generated by primary fields in one-to-one
correspondance with those on the right hand side of the (Verlinde)
fusion rules $\phi_a\phi_b\rightarrow N_{ab}^c \phi_c$.

The $A_k$ $\CN=2$ minimal model can be obtained as a deformation of the
$SU(2)_k$ WZW model, and its primary fields
$\phi^l_m$ are labelled similarly, by two integers $0\le l\le k$ (the $SU(2)$
representation label) and $0\le m<2k+4$ (the charge under the
$U(1)$ of $\CN=2$) up to a $\BZ_2$ identification $(l,m)\cong (k-l,m+k+2)$.
The fusion rules are the product of $U(1)$ fusion rules (i.e. $\BZ_{k+2}$
charge conservation)
with $SU(2)_k$ fusion rules.

Before implementing the GSO projection,
the Gepner model boundary conditions are labelled by a set of such integers,
and are all A boundary states (since they correspond to left-right symmetric
fields).
The GSO projection then
restricts the closed string spectrum
to (odd) integer total $U(1)$ charge $\sum m$, while twisted
states with $m_L\ne m_R$ are added.  The restriction has the effect of reducing
the number of distinct A boundary states, while the twisted sectors provide new
candidate B boundary states.

The final result for the $(3)^5$ model is that all boundary states are labelled
by
a set of five $L_i \in \{0,1\}$;
the A boundary states are also
labelled by five $M_i$ satisfying one relation
and form representations
of $\BZ_5^4\times S_5$
discrete symmetry, while the B boundary states have a single $M$
label and represent a $\BZ_5$ discrete symmetry.
These are the known discrete symmetries of the CFT at the Gepner point;
it is known to be equivalent to the Fermat
quintic $\sum_{i=1}^5 Z_i^5  = 0$ in $\BP^4$
with manifest $\BZ_5^4$ symmetry, at
a special point in K\"ahler moduli space with quantum $\BZ_5$ symmetry.

The modified geometric hypothesis of section 2 would imply that these
A branes are
exactly the {\sLm}s of the Fermat quintic and we can test this idea for the
known {\sLm}s.  These are obtained by taking a real section
$\Im e^{2\pi i m_i/5} Z_i = 0$: topologically these are
$\BR\BP^3$'s, which fall into the same representation of
$\BZ_5^4\times S_5$ as two sets of boundary states: those with all $L_i=0$ and
those with all $L_i=1$.  How can we tell which (if either) is their
counterpart?

A strong check of any proposed identification is that the geometric
intersection
number of a pair of $3$-branes must agree with the quantity 
$\Tr_{ab} (-1)^F$ in this
sector of the open string theory. \cite{df}  This can be seen by considering
electric-magnetic charge quantization in the resulting $d=4$ theory.  This
computation is a special case of those in \cite{rs}\ and it turns out the
$L_i=1$ states match this intersection form, while
the $L_i=0$ states do not (they presumably correspond to some
other {\sLm}s).
So far this is in agreement with both the original and the modified
geometric hypothesis.

However, one also finds that the $L_i=1$ brane world-volumes
have a massless chiral multiplet, and this disagrees with the geometric
prediction of \cite{mclean}.
As discussed in \cite{bdlr}\ it is likely that this is lifted by a
superpotential,
but even so this contradicts the strongest form of the geometric hypothesis, in
which both this massless field and the superpotential would have matched.
It does not contradict the modified geometric hypothesis, which allows the
A brane superpotential to depend on the K\"ahler form,
and furthermore shows that massive fields in the large volume
limit can come down to become (linearized) moduli.
Such effects and even jumping of the dimension of moduli space
are known to be possible in the B picture; perhaps this superpotential would be
manifest in a mirror description.

Turning to the B branes, we have more intuition for which of these exist in
the large volume limit: namely the condition (\ref{stability}) must be
satisfied
(if $Q_6\ne 0$; there is an analogous statement if $Q_6=0$ but $Q_4\ne 0$).
Although bundles on the quintic are by no means classified, various
considerations
suggest that generic charge
vectors for which the discriminant (the left hand side of (\ref{stability}))
is sufficiently large will be associated to stable bundles.

Thus it is interesting to express the charges of the B boundary states in large
volume
terms, and compare.  A precise form of this comparison is to choose a path in
K\"ahler moduli space from the Gepner point to the large volume limit, and use
the flat $Sp(2r,\BZ)$ connection provided by special geometry to transport the
charge
lattices between the two regimes.

The K\"ahler moduli space and prepotential for the quintic is of course well
known from the famous work of Candelas et. al. \cite{candelas} which computed
the periods of the three-form on the mirror.
To review the structure of this moduli space: it is a Riemann sphere with three
singularities, a large volume limit at $z\rightarrow \infty$, the Gepner point
with
a $\BZ_5$ quotient singularity at $z=0$, and finally a ``conifold'' singularity
at
$z=1$ at which a three-cycle of the mirror degenerates (has zero period).
It turns out \cite{candelas,greenekanter} that on the original quintic this is
precisely the central charge of the ``pure'' (trivial gauge field) six-brane.
The periods $\Pi_i(z)$ can be
obtained as solutions of Picard-Fuchs ODE's or more explicitly as series
expansions
around each singularity, with radius of convergence determined by the locations
of the others.

Two concrete results are now needed from this analysis.
First, the mirror map gives us an
appropriate basis for the large volume limit -- central charges of the
individual
$2p$-branes.
Second, given that the central charge of a brane with charge vector $Q^i$
is $Z=Q^i \Pi_i(z)$, the transition functions of the flat connection on the
charge
lattice are simply the linear transformations between different bases
$\Pi_i(z)$
adapted to different regions of moduli space (these are connection formulas
for generalized hypergeometric functions).
This tells us what the $2p$-brane central charges will be at the Gepner point.

In principle these could already be compared with a precise computation of the
central
charges of our boundary states, but such a comparison will run into tricky
problems of normalization.
The best way to study the charges of D-branes -- as was done in the very first
example \cite{pol} -- is to instead compute the interaction between two
D-branes
in the open string channel, as this is canonically normalized (it is a
partition function).
Indeed the simplest quantity of this type is the intersection form $\Tr_{ab}
(-1)^F$
discussed above and thus the simplest way to proceed is to express the known
large volume intersection form in terms of a natural basis at the Gepner point
(one which represents the quantum $\BZ_5$ symmetry in a simple way) and
compare this intersection form with the intersection forms of the boundary
states.

It turns out that the resulting boundary state charges are simple when
expressed
using the basis first postulated by Candelas -- the zero-brane period and its
$\BZ_5$ images.  The states of minimal charge are the five $L_i=0$ states;
one of these turns out to be the pure six-brane
$\bra{6B} \equiv (\matrix{1&0&0&0})$, and to
get the others we just need to know the $\BZ_5$ monodromy in the large volume
basis, which is given in \cite{candelas}.  In the conventions of 
\cite{bdlr}
it is\footnote{
Note that these are conventions in which the charge vectors include
the factor $\sqrt{\hat A}$, which are not the conventions of (\ref{stability}).
The latter are also given in \cite{bdlr};
they are the ones in which the large volume monodromy is simple
but charges are not necessarily integral.}
\eqn\label{eq:z5mono}
g \equiv (\matrix{Q_6&Q_4&Q_2&Q_0})
\rightarrow (\matrix{Q_6&Q_4&Q_2&Q_0})
\left(\matrix{-4&-1&-8&5\cr
-3&1&5&3\cr
1&0&-1&1\cr
-1&0&0& 1}\right)
\enq
and thus the others are $\bra{6B}g^M$.
The charges for states with $L=\sum L_i>0$
can be derived from these by using
the fusion rules: essentially, they are $\bra{Q_6} (1+g)^L g^M$.

One surprise of the result is that the D$0$-brane
is not present (as a rational
boundary state; this is not to say that it does not exist at the Gepner point).
It appears that this is also consistent with the geometric
hypothesis in the following sense:
any location we might pick for the D$0$ would break some of $\BZ_5^4$, but
all of the rational B boundary states are singlets under $\BZ_5^4$, so we
should not
find the D$0$ in this analysis.\footnote{
It appears that other Gepner models can contain the D$0$ as a boundary
state.\cite{dr}}

Looking at the charges of all of the boundary states,
they appear to be consistent with the original geometric hypothesis,
at least in the weak sense that they are all consistent with
(\ref{stability}).  Not too much more is known about vector bundles
on the quintic so it is hard to be more precise.

On the other hand, the monodromy (\ref{eq:z5mono}) in general can take
solutions of (\ref{stability}) into non-solutions, making it highly
implausible that it is a symmetry of the entire brane spectrum.
This is reminiscent of related phenomena in the study of $\CN=2$
gauge theory, and we turn to this analogy.

\section{Marginal stability on the quintic}

As we saw in the previous section, the D$0$-brane is not a rational
boundary state for the Gepner quintic.  This leads one to wonder whether
it exists in the stringy regime at all, and more generally how much
the spectrum of branes varies as we move around.

In generic $\CN=2$, $d=4$ theories, the spectrum of BPS states depends
on the moduli, but it varies in a highly constrained way.  A state of
charge $Q$ will generically be stable under variations of the moduli,
but there exist can lines of marginal stability (or ``jumping lines''),
on which the state can decay to BPS states of charge $Q_1$ and $Q_2$,
if the condition
\eqn
|Z(Q)| = |Z(Q_1)| + |Z(Q_2)|
\enq
is satisfied.  Here $Z(Q)=Q\cdot\Pi(z)$ is the central charge in terms of
a vector of periods $\Pi(z)$ at a point $z$ in moduli space; for the A
branes these are the periods of the three-form $\Pi=\int\Omega$
(normalized to $\int \Omega\wedge\bar\Omega =1$).

The most familiar examples are supersymmetric gauge theories, which have
been studied in great detail.  For example, pure $SU(2)$ $\CN=2$ gauge theory
(the original Seiberg-Witten solution) has a line of marginal stability
which goes through the massless monopole and dyon points and separates the
strong and weak coupling limits.  The strong coupling BPS spectrum consists
only of the monopole and dyon,
the two states responsible for the singularities.
This phenomenon was necessary
as otherwise monodromies around the massless monopole point would produce
states with arbitrarily large electric charge, which are not present
in the known semiclassical spectrum.

Besides the known semiclassical spectrum, a number of constraints follow
from the solution for the prepotential and justify this result.
The primary constraint is the physical correspondance between singularities
and massless states \cite{strom}:
if $Z(Q)$ vanishes at some $z$, either there is a
corresponding singularity which we can think of as coming from integrating
out this state at nearby points, or else the state must not exist at $z$.
If it exists at some $z'$, there must be a line of marginal stability
separating $z$ and $z'$.
This is quite strong as it turns out that the ratio of the two periods
$a_D/a$ assumes all possible real values (in all the asymptotically free
$SU(2)$ theories in fact) and thus every charge can be constrained.
One sees this most easily by combining
the result (easily verified numerically)
that $\Im a_D/a$ changes sign between weak and strong coupling regimes
with the $SL(2,\BZ)$ transformation properties of $a_D/a$ (which force
the line $\Im a_D/a=0$ to connect the massless monopole and dyon points).

Our earlier observation that
the $\BZ_5$ monodromy obtained by encircling the Gepner point in the quintic
does not fit well with the known constraints on the large volume spectrum
is our first suggestion that similar phenomena will obtain here.
There is also a qualitative similarity to the change of sign of $\Im a_D/a$.
Let the conifold point be $z_c=1$: here the six-brane becomes massless,
$\Pi_6\sim z-z_c$.  Although the other periods are not analytic here,
they are still continuous: $\Pi \sim (z-z_c) \log (z-z_c) + {\rm regular}$.
Thus just as in gauge theory,
$\Im \Pi_6/\Pi_0$ changes sign as we go through this point.

This starts to suggest that the gauge theory picture with its drastic
change in the spectrum might also be possible here.  Unfortunately few of
the other elements of the story there have been developed for the quintic
(or indeed any CY) moduli space.  In particular, the appropriate analogs
of $SL(2,\BZ)$ and the fundamental region are not known, making it difficult
to get a good global picture of the moduli space.

The boundary state results show us that the answer will not be as simple
as that for gauge theory -- the spectrum will not collapse simply to the
states which can become massless.  We should also not assume that
all of the boundary states exist at large volume.

To study this one can simply follow all of the central charges for boundary
states out from the Gepner point to the large volume limit, to see what
happens.  One expects more marginal stability lines in the neighborhood
of the conifold point, so to minimize the possibilities for decay
we choose the trajectory $z$ real and negative
opposite to it in moduli space.  We then numerically integrated the
Picard-Fuchs equations (and checked the results against the series
expansions of \cite{candelas}) to get the periods and thus the BPS masses.

Using these to compute the masses of BPS branes with the charges of
all rational boundary states produces a surprise: one of them has its period
go through zero! In other words, there exists a BPS state at the Gepner
point whose mass appears to go to zero at a non-singular point $X$ in moduli
space.
(Readers who want proof that this is not an error of numerics or
conventions will find a semi-qualitative proof in the appendix.)

This in itself is not inconsistent as long as there exists a line of
marginal stability separating the point $X$ from the Gepner point.
At this point we run into one of the main difficulties in studying
these questions for CY:
there are an infinite number of candidate marginal stability lines, and
we need more knowledge about the BPS spectrum to decide which are real
(i.e. the decay takes place, which requires the
states of charge $Q_1$ and $Q_2$ to actually exist on the line).
This is closely related to the fact that at a generic point in moduli
space, there exist charges $Q$ such that $|Z(Q)|<\epsilon$ for any
positive $\epsilon$, no matter how small.  Consider the Gepner
point: there the periods are the fifth roots of unity, so the set of $Z(Q)$
is a $Z_5$ symmetric lattice embedded in the complex plane.

Although we have not as yet found the true marginal stability lines,
we can at least try to postulate a pair of charges
$Q_1$ and $Q_2$ into which the problematic state can decay and whose masses
do not cross zero on the way to large volume.  This is not hard to do,
and thus the existence of such a marginal stability line
seems perfectly plausible -- there seems no reason to doubt the consistency
of the theory.

Thus we have proof of the existence of at least one marginal stability
line; given that we have two points at which $Z(Q)$ vanishes for ``simple''
charges
$Q$ it is quite likely that many other true marginal stability lines pass
through these points.

An even stronger consideration of this type is to follow large volume
branes to the Gepner point: it is easy to find charge vectors
satisfying (\ref{stability}) whose period goes through zero on this axis.
If it is true that these generally correspond to stable bundles,
we have many more examples.

All this starts to be significant evidence
for the claim that the BPS spectrum is rather different in the stringy regime.

\subsection{A note on attractor points}

A question related to marginal stability but somewhat simpler has arisen in the
study of BPS black holes in CY compactification.  It has been shown \cite{fks}
that the
entropy of such black holes is governed by the ``attractor mechanism.''
Given a black hole of large charge $Q$,
the consistency condition for a covariantly constant spinor is a
first order equation which is just gradient flow on the moduli space to a
minimum of
the quantity $S(z)=|Q\cdot \Pi(z)|$; the entropy is the minimal value
$S_{min}(Q)$.

For some $Q$, it is possible that
$S_{min}(Q)=0$.  In the previously known examples (such as the state which goes
massless at the conifold point), the state existed at the minimizing point in
moduli
space and produced a singularity in the moduli space metric, modifying the
discussion.
What we have found here is a $Q$ for which the discussion above leads to a
contradiction
(as noted in \cite{moore}) --
the attractor equation breaks down (has no sensible solution)
before reaching the horizon, so this is not a failure of supergravity.
Indeed, this could be interpreted as an argument that such black holes cannot
exist, and an observation consistent with this idea is that
(at least in some cases) the condition $S_{min}(Q)=0$ reduces to
the negation of (\ref{stability}) in the large volume limit on the quintic,

However, we have found a particle with (small) charge
$Q$ and $S_{min}(Q)=0$
at the Gepner point, so we have a paradox.  We can take $N$ of these particles and
put them into a small region of space, using only total energy $Nm+\epsilon$.
For $N$ sufficiently large, one would certainly expect that they form a black
hole $NQ$,
for which the previous argument applies.

What is going on ?  The resolution will almost certainly use the fact that --
as a single
brane -- the object in question was unstable at the minimizing point.  One
scenario
is that the final stable object is a bound state of two black holes of charges
$NQ_1$ and $NQ_2$ with a hard core repulsive potential. \cite{dougmald}
This would evade the previously
cited argument, which assumed a spherically symmetric configuration.

It seems likely that more surprises along this line await us.

\section{Conclusions and further directions}

D-branes have played a central role in the study of superstring and M theory
duality.
Quite a lot has been understood about compactifications with enhanced
supersymmetry,
but eventually we will need to deal with the physical cases of $\CN=0$ (and
hopefully
$\CN=1$!) supersymmetry in four dimensions.

A large class of $\CN=1$ supersymmetric string compactifications can be
obtained by using
D-branes on Calabi-Yaus.  Many of these are related to known constructions
(F theory or the strong coupling limit of heterotic strings) but what
I have tried to show here is that we can make further progress by using special
properties
of the weak type \II\ string coupling limit, namely the close relation between
D-brane
theories which fill different parts of Minkowski space (e.g. D$3$ and
D$0$-branes), and
the powerful tools of mirror symmetry and exactly solvable CFT.

A reasonable goal for the current work is to settle the geometric hypothesis
and
modified geometric hypothesis as described here -- namely, to show that the
superpotential and D terms depend only on complex and K\"ahler data for B
branes
(the reverse for A branes) and answer the following questions:
\begin{enumerate}
\item Are all A branes $3$-branes wrapped on \sLm's, even for a stringy CY ?
Are all marginal stability lines and decays described by the local
intercommutation of $3$-branes ?

\item If so, do the mirror symmetry predictions for the spectrum of B branes
agree with geometric predictions at large volume?  What does the semistability
condition translate into in the A picture ?

\item Is the spectrum of B branes on stringy CY's very different from
the large volume spectrum (as the results here suggest) ?
If so, is it finite or perhaps characterized by simple inequalities
analogous to (\ref{stability}) ?

\item Is knowledge of the large volume spectrum and the exact prepotential
enough to determine the spectrum throughout moduli space (using consistency
arguments of the sort which worked for supersymmetric gauge theory) ?

\item Can we make a complete statement about the potential and moduli space on
these branes (presumably combining B picture information to get the
superpotential
and A picture to get the D terms) ?

\item Can we extend this picture to finite string coupling, perhaps by making
contact
with the heterotic string limits of the same models ?
\end{enumerate}

A longer term goal will be to understand terms in the effective action which
are not
so strongly constrained by supersymmetry, such as the D$0$-metric on a CY.

Perhaps interesting non-supersymmetric models can be obtained by considering
non-BPS
space-filling brane configurations, along the lines of \cite{sen,kachru}.

\section*{Acknowledgments}

I would like to thank T. Banks, I. Brunner, D.-E. Diaconescu,
B. R. Greene, M. Kontsevich, S. Kachru, A. Lawrence, J. Maldacena,
G. Moore, C. Romelsburger, C. Vafa and E. Witten for collaboration and helpful
discussions.

\noindent
This research is supported in part by DOE grant DE-FG05-90ER40559.

\section*{Appendix}

We give a semi-qualitative argument for the vanishing period, using the
results for the periods of the mirror to the quintic in \cite{candelas}.
They are functions of a complex modulus $\psi$ which covers the Riemann sphere
with
three punctures. $\psi\rightarrow\infty$ is the large volume limit, with
$t = B + iV = -{5\over 2\pi i}\log(5\psi)$.
$\psi=1$ is the conifold point, and $\psi=0$ is the
Gepner point.
$\psi\rightarrow \alpha\psi$ with $\alpha=e^{2\pi i/5}$ is the $\BZ_5$ quantum
symmetry
of the Gepner point -- it leads to the same bulk theory but acts
as an $Sp(4,\BZ)$ monodromy on the brane spectrum.

Candelas et al. use a basis $\omega_k(\psi)$ where $\omega_0$ is the $0$-brane
period in the
large volume limit and the others are its images under the $\BZ_5$.
of the Gepner point.  These are multi-valued on the $\psi$ plane and thus
it is necessary to take care with the domains of definition.

There are three lines along which the periods have simple
reality properties.  We define the line $A$ to be
$\psi=x$ real with $x >1$, the line $B$ as $\psi=x$ real
satisfying $0\le x<1$, and the line $C$ as
$\psi=e^{2\pi i/10} x$ with $x$ real and positive.

{}From the explicit series expansions
for the periods it is easy to check the following qualitative properties:
\begin{enumerate}
\item
Near the Gepner point,
$\omega_j(\psi) \rightarrow -\alpha^{2+j} C \psi$
with $C=\Gamma(1/5)/\Gamma(4/5)^4$ a positive real constant.

\item In the large volume limit,
$
\omega_j(\psi) \sim {S_{j3}\over 6} t^3
$
where $S_{j3} = 0, 5, -15, 15, -5$ for $j=0,1,2,3,4$.

\item
Along $B$ we have
$(\omega_j(x))^* = \omega_{1-j}(x)$
and along $C$ we have
$(\omega_j(\alpha^{1/2} x))^* = \omega_{-j}(\alpha^{1/2} x)$
(this must be checked using both large and small volume expansions).
\end{enumerate}

{}From \cite{bdlr}, one can check that the period
\eqn
\Pi_X = \omega_1 - \omega_4
\enq
is the central charge of a B boundary state $L_1=1$, $L_i=0$ for $i>1$.

We now argue that $\Pi_X$ will have a zero along the axis C.
{}From (iii) we see that $\Pi_X$ is purely imaginary along this
axis, so if the imaginary part changes sign between the Gepner and large
volume limits it must have a zero.  This can be checked explicitly given the
limiting behaviors we quoted.

A way to see that this was inevitable is to consider the behavior of
the six-brane period $\Pi_6=\omega_1-\omega_0$ on the loop $ABC$ in moduli space.
At large volume,
$
\Pi_6 \sim_{\psi\sim\alpha^{1/2}\infty} {5\over 6} t^3
\sim {5\over 6} -iV^3
$
so it comes in from negative imaginary infinity towards zero.
Along A and B $\Pi_6$ is purely imaginary and as we know it crosses
zero at $\psi=1$ (the conifold point) and comes out the other side,
to reach its value at the Gepner point
$
\Pi_6(\psi) \sim_{\psi\sim 0} C (\alpha^2 - \alpha^3) \psi
= C 2i \sin {\pi\over 5} \psi.
$
As we come back along the axis C, we know that the six-brane does not
become massless anywhere, so $\Pi_6$ must move off into the complex
plane to avoid the origin, finally joining the same asymptotics
$\Pi_6 \sim -5/6 iV^3$ we had at large positive $\psi$.

This behavior implies that $\Pi_6$ must cross the real axis at some
point, and since $\omega_0$ is real all along C, $\omega_1$ must become
real at this point.


\bigskip
\section*{References}
\begin{thebibliography}{100}

\bibitem{trans}
{\tt http://www.aei-potsdam.mpg.de/cgi-bin/viewit.cgi}\\ and
{\tt http://www.physics.upenn.edu/particle\_meeting/talk\_titles.html}.
\bibitem{bbs} K.~Becker, M.~Becker and A.~Strominger,
\NP\ {\bf B456}, 130 (1995) hep-th/9507158.
\bibitem{bdl}M.~Berkooz, M.R.~Douglas and R.G.~Leigh,
\NP\ {\bf B480}, 265 (1996), hep-th/9606139.
\bibitem{bcov} M. Bershadsky, S. Cecotti, H. Ooguri and C. Vafa,
\NP\ {\bf B405}, 279 (1993).
\bibitem{bdlr}
I. Brunner, M. R. Douglas, A. Lawrence and C. R\"omelsberger, hep-th/9906200.
\bibitem{calmal} C. Callan and J. Maldacena,
\NP\ B513 (1998) 198-212; hep-th/9708147.
\bibitem{candelas} P.~Candelas, X.C.~De La Ossa, P.S.~Green and L.~Parkes,
\NP\ {\bf B359}, 21 (1991).
\bibitem{fract} D.-E. Diaconescu, M. R. Douglas and J. Gomis,
 JHEP 9802 (1998) 013; hep-th/9712230.
\bibitem{dg} D.-E. Diaconescu and J. Gomis, hep-th/9906242.
\bibitem{dr} D.-E. Diaconescu and C. R\"omelsburger, to appear.
\bibitem{doug}  M. R. Douglas, hep-th/9901146.
\bibitem{dm} M. R. Douglas and G. Moore, hep-th/9603167; \\
E. Gimon and J. Polchinski, hep-th/9601038.
\bibitem{dgm} M. R. Douglas, B. R. Greene and D. R. Morrison, \\
\NP\ B506 (1997) 84-106; hep-th/9704151.
\bibitem{douggreene} M. R. Douglas and B. R. Greene,
Adv.Theor.Math.Phys. 1 (1998) 184; hep-th/9707214.
\bibitem{df} M. R. Douglas and B. Fiol, hep-th/9903031.
\bibitem{dougmald} M. R. Douglas and J. Maldacena, work in progress.
\bibitem{dta} M. R. Douglas, to appear.
\bibitem{fks} S. Ferrara, R. Kallosh and A. Strominger,
\PR\ D52 (1995) 5412-5416, hep-th/9508072.
\bibitem{govind} S. Govindarajan, T. Jayaraman and T. Sarkar, hep-th/9907131.
\bibitem{gretc} B.R. Greene,
\NP\ B525 (1998) 284-296; hep-th/9711124;\\
T. Muto, \NP\ B521 (1998) 183-201; hep-th/9711090.
\bibitem{greenekanter} B.R. Greene and Y. Kanter,
\NP\ {\bf B497}, 127 (1997) hep-th/9612181.
\bibitem{grossruan} M. Gross, alg-geom/9710006, math.AG/9809072; \\
math.AG/9909015; W.-D. Ruan, math.DG/9904012 and math.DG/9909126.
\bibitem{gutperle} M. Gutperle and Y. Satoh, hep-th/9808080 and hep-th/9902120.
\bibitem{harvey} F. Reese Harvey, {\it Spinors and Calibrations}, Academic
Press,
1990.
\bibitem{hm} J.A. Harvey and G. Moore,
Comm. Math. Phys. {\bf 197} 489, (1998) hep-th/9609017.
\bibitem{joyce} D. Joyce, hep-th/9907013.
\bibitem{kachru} S. Kachru and J. McGreevy, hep-th/9908135.
\bibitem{ks} S. Kachru and E. Silverstein, Phys.Rev.Lett. 80 (1998) 4855-4858;
 hep-th/9802183.
\bibitem{lnv} A. Lawrence, N. Nekrasov and C. Vafa,
\NP\ B533 (1998) 199-209; hep-th/9803015.
\bibitem{mclean} R. C. McLean, Ph. D. thesis, Duke 1990; N. Hitchin,
dg-ga/9711002.
\bibitem{moore} G. Moore, hep-th/9807087.
\bibitem{ooy} H. Ooguri, Y. Oz and Z. Yin,
\NP\ B477 (1996) 407; hep-th/9606112.
\bibitem{pol} J. Polchinski, \PRL\ 75 (1995) 4724-4727; hep-th/9510017.
\bibitem{rs} A. Recknagel and V. Schomerus,
\NP\ {\bf B531}, 185 (1998), hep-th/9712186.
\bibitem{sardoinfirri} A. V. Sardo Infirri, alg-geom/9610004.
\bibitem{sen} A. Sen, hep-th/9904207.
\bibitem{sharpe} E. Sharpe,
Adv.Theor.Math.Phys. 2 (1999) 1441-1462, hep-th/9810064.
\bibitem{strom} A. Strominger,
\NP\ B451 (1995) 96-108; hep-th/9504090.
\bibitem{syz} A. Strominger, S.-T. Yau and E. Zaslow,
\NP\ B479 (1996) 243; hep-th/9606040.
\bibitem{vafa} C. Vafa,  hep-th/9804131.

\end{thebibliography}
\end{document}